\documentclass[conference]{IEEEtran}

\usepackage{graphicx} 
\usepackage{booktabs} 
\usepackage[tbtags]{amsmath}
\usepackage{amsfonts}
\usepackage{amssymb}
\usepackage{dblfloatfix}
\usepackage[caption=false,font=footnotesize]{subfig}
\usepackage{color}
\usepackage{enumitem}
\usepackage{amsthm}
\usepackage{algorithmic}
\usepackage{multirow}
\usepackage{array}
\usepackage{cite}
\usepackage[normalem]{ulem}
\usepackage{bbm}
\usepackage{mathtools}
\DeclarePairedDelimiter\myfloor{\lfloor}{\rfloor}

\begin{document}

\title{Repeat-Authenticate Scheme for Multicasting of Blockchain Information in IoT Systems}

\author{\IEEEauthorblockN{Pietro Danzi, Anders E. Kal{\o}r, \v{C}edomir Stefanovi\'c, Petar Popovski,\\}
\IEEEauthorblockA{Department of Electronic Systems, Aalborg University, Denmark \\
Email: \{pid,aek,cs,petarp\}@es.aau.dk }
}

\maketitle

\begin{abstract}
We study the problem of efficiently disseminating authenticated blockchain information from blockchain nodes (servers) to Internet of Things (IoT) devices, through a wireless base station (BS).
In existing blockchain protocols, upon generation of a new block, each IoT device receives a copy of the block header, authenticated via digital signature by one or more trusted servers. Since it relies on unicast transmissions, the required communication resources grow linearly with the number of IoT devices.
We propose a more efficient scheme, in which a single copy of each block header is multicasted, together with the signatures of servers.
In addition, if IoT devices tolerate a delay, we exploit the blockchain structure to amortize the authentication in time, by transmitting only a subset of signature in each block period.
Finally, the BS sends redundant information, via a repetition code, to deal with the unreliable wireless channel, with the aim of decreasing the amount of feedback required from IoT devices.
Our analysis shows the trade-off between timely authentication of blocks and reliability of the communication, depending on the packet loss rate offered by the channel.
The numerical results show that the performance benefits of the proposed scheme makes it a viable starting point for designing new lightweight protocols for blockchains.
\end{abstract}

\section{Introduction}

In recent years, there has been a growing interest in understanding the potential applications of blockchains to the Internet of Things (IoT) landscape~\cite{christidis2016blockchains}.
However, the literature on the integration of the blockchain protocols with the wireless networks is still scarce~\cite{danzi2019communication}.
In this respect, discovering the trade-offs and potential bottlenecks of blockchain protocols plays a pivotal role in the context of IoT systems, in which devices typically have constrained energy resources and are connected by low-cost wireless networks, e.g. Wi-Fi, Bluetooth, or LoRaWAN \cite{durand2018resilient}.

In this work, we investigate the communication cost incurred by a wireless base station (BS) for sending blockchain information to IoT devices.
In legacy blockchain synchronization schemes, a device, connected to the blockchain network via the BS, receives an update whenever a new block is generated by the blockchain network.
Clearly, such schemes do not scale well when the number of devices connected to the same BS grows, leading potentially to  a communication bottleneck. A related problem is that the legacy schemes use reliable transport layer (e.g. TCP), which involves a significant signalling overhead and is thus unsuitable for low-energy IoT devices.

Our proposal is based on the key observation that if devices are connected to the same blockchain network, they are highly likely to be interested in receiving the same information, namely the same blocks or, in case of lightweight clients~\cite{danzi2018delay}, just the block headers.
The only difference among the devices is the set of servers that each device trusts. Hence, the block should be authenticated by several servers in such a way that each node trusts at least one of the authentication servers.\footnote{The servers authenticate the data packet carrying the block and should not be confused with the blockchain validators, which sign the block.}
The servers of the blockchain network provide authentication of blocks by sending to devices a digital signature that signs the block.
Therefore, in our proposal the end-to-end channel from blockchain nodes to IoT devices is authentic but not confidential. Still, a confidential channel can be set up when a device needs to exchange additional information through transactions.
\begin{figure*}[tb]
\centering
\subfloat[]{ \includegraphics[width=0.65\columnwidth]{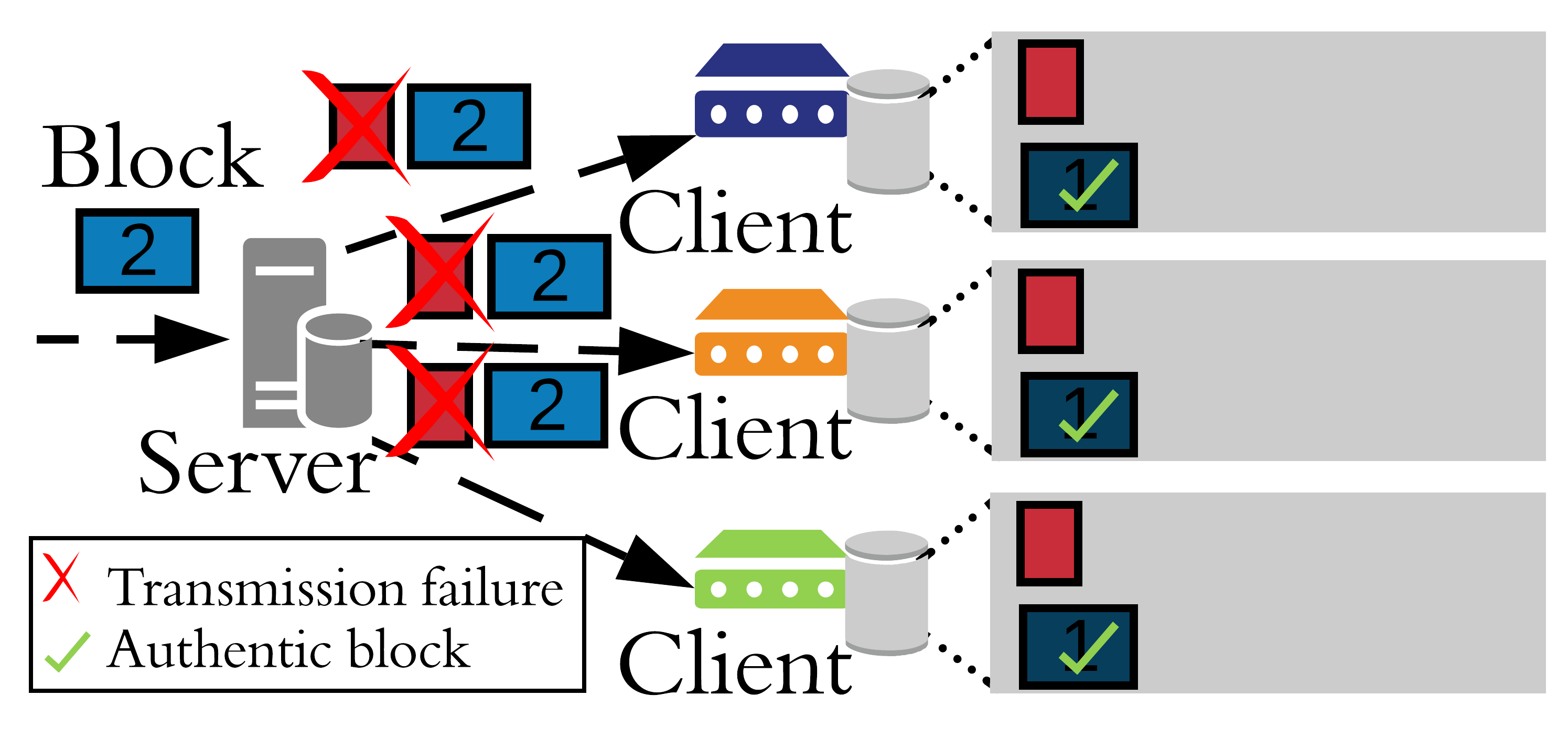}}
\subfloat[]{ \includegraphics[width=0.65\columnwidth]{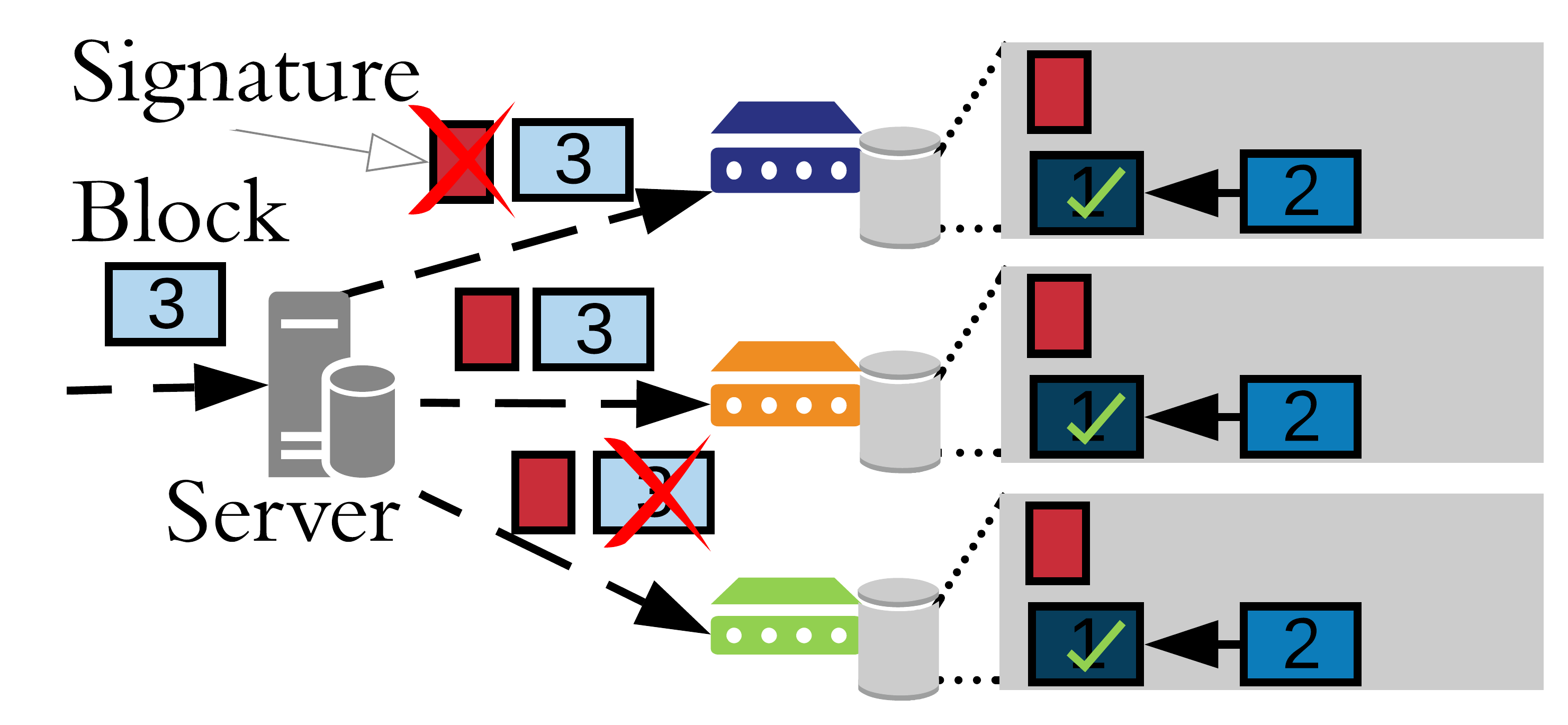}}
\subfloat[]{ \includegraphics[width=0.65\columnwidth]{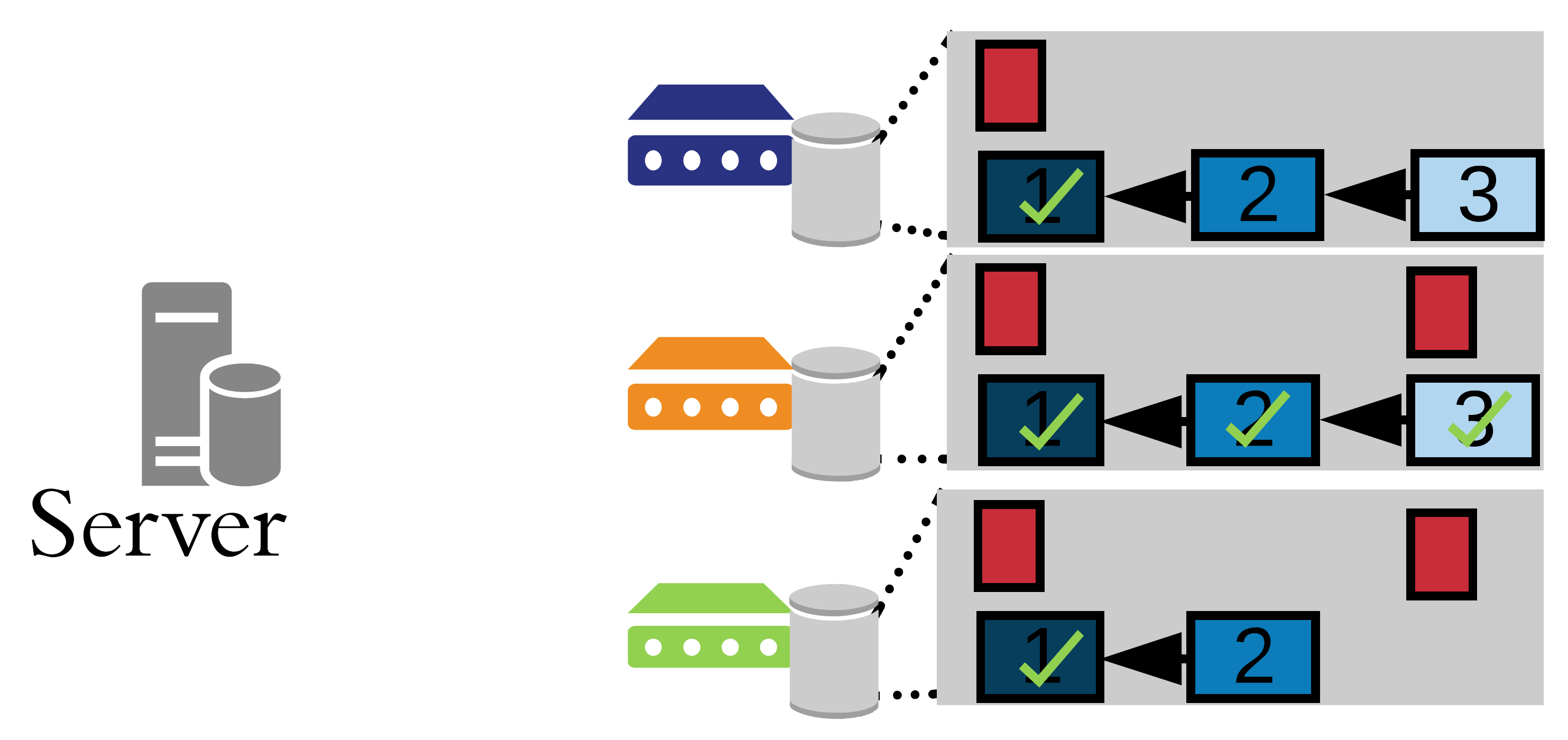}}
    \caption{Representations of authenticated blockchain for three alternative cases (a) before and (b)-(c) after the transmission of two authenticated blocks.}
    
 \label{fig:amm}
\end{figure*}

By leveraging the broadcast nature of the wireless medium, we design a scheme in which the BS multicasts blocks to all devices, together with a set of server signatures. Reliable transmission is, in general, provided by forward error correction (FEC). In this work we attain reliability in the simplest form of FEC, which is a mere repetition of the blocks.
The signatures are not repeated, since we rely on the \emph{signature amortization} property that is inherent to the blockchain protocols and allows a device to authenticate all previously chained blocks by receiving a single signature.
Depending on the progress of the reception of blocks and signatures, several cases may arise, as illustrated in the examples in Fig.~\ref{fig:amm}, in which there is a server and three clients, which are initially synchronized. The server first multicasts a block to the clients, but none of the clients receive the signature, see Fig.~\ref{fig:amm}(a). Consequently, since the clients cannot verify the block, they are now delayed by one block, see Fig.~\ref{fig:amm}(b). The server then sends a third block to the three clients. The first client receives the block without signature, increasing its authentication delay to two blocks, see Fig.~\ref{fig:amm}(c).
For the second client, both the third block and its signature are received, making the second block authenticated as well, thanks to the chaining.
The third client only receives the signature of the third block.
In this case, the number of non-authenticated blocks is also two, as the signature of the third block can not be chained because signatures are chained through blocks. Hence, the use of amortized signatures mitigates the effect of packet loss. As soon as the blocks are chained, the client can authenticate its copy of the blockchain using only a subset of the signatures.

Through analysis of the scheme, we show that the BS can trade-off the timely authentication of block streams, and their reliable transmission, by selecting the length of the code, i.e. repetitions of the blocks.

The remainder of the paper is organized as follows.
Section~\ref{sec:background} provides a brief introduction to blockchain and to related works.
Section~\ref{sec:system} presents the system model.
Section~\ref{sec:schemes} elaborates the proposed scheme, while Section~\ref{sec:analys} is devoted to its analysis. Section~\ref{sec:results} presents the evaluation and Section~\ref{sec:conclusion} concludes the paper.

\section{Background}\label{sec:background}
\subsection{Blockchains}

\begin{figure}[!tb]
\centering
  {\includegraphics[width=0.43\columnwidth]{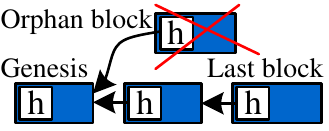}}
    \caption{Representations of blockchain. $\text{`h'}$ indicates the hash value of the previous block.}
 \label{fig:fork}
\end{figure}

In a blockchain network, a (possibly large) set of nodes store a copy of a same ledger.
The ledger records the modifications to the state of a system, e.g. a financial accounting~\cite{nakamoto2008bitcoin}.
The modifications are batched into information blocks, which also include a header that contains metadata.
The initial state is described in a block called genesis, see Fig.~\ref{fig:fork}.
Upon each modification of the state, a new block is propagated through the network, becomes locally validated by the nodes, and, if valid, is appended to each nodes' local copy of the ledger.
Each block contains in their header the hash value of the previous block header, which enforces an order and ensures uniqueness of the ledger, see Fig.~\ref{fig:fork}.
To avoid uncontrolled generation of blocks, nodes can only propagate blocks that fulfil certain consensus rules, e.g. proof-of-work \cite{nakamoto2008bitcoin}.
This provides consistency of the ledgers, without a centralized trusted authority.

The propagation delays in the network might cause forks, i.e. presence of conflicting, valid chains of blocks \cite{decker2013information}.
The forks are resolved by voting on which of the co-existing chains should be considered valid~\cite{nakamoto2008bitcoin}.
After the resolution of a fork, the block that remains without children, see Fig.~\ref{fig:fork}, termed \emph{orphan} in Bitcoin, is deleted from memory.
In principle, a fork can last many blocks, however regular blockchain networks are well connected to avoid generating consecutive orphan blocks.

As the validation of blocks is an expensive process in terms of memory and computing resources, lightweight clients typically request only the block header, instead of the entire block, delegating the verification to trusted nodes.
In addition, they may request part of the ledger that they are interested in observing~\cite{danzi2017analysis, danzi2018delay}. Since this entails full trust in the nodes, in order to detect misbehaving ones, a lightweight client can request the information from a set of nodes, and verify that the received block headers are consistent.\footnote{Alternatively, they can implement incentive-based protocols, e.g. see~\cite{gruberunifying}, which are not treated in this paper.}
The model described in the following text focuses on lightweight clients, which represent typical IoT devices.

\subsection{Related work}

The problem of reliable and authenticated multicasting of data streams is a well-investigated subject~\cite{nafaa2008forward}.
Here, a message is authenticated when it is digitally signed by a trusted party and the signature is received. However, message and signature are not necessarily sent together.
Hence, when a message is received, but not authenticated, it is considered useless.

In systems where authentication delay is tolerated, amortized signatures, based on cryptographic hash functions~\cite{merkle1987digital}, can be used to reduce the communication cost \cite{park2003efficient}.
The idea is to include in each message the hash value of the previous message, thus \emph{chaining} them in a sequence.
In this way, the reception of a valid signature for a message makes the previous messages authenticated as well.
This feature, illustrated in Fig.~\ref{fig:amm}, is already embedded in the blockchain by design, thanks to the block chaining.

When a delay is not acceptable, the signatures of different parties can be combined to a shorter one, generated by aggregate signature schemes~\cite{boneh2018compact}.
When the aggregate signature is valid, it means that all servers authenticated it.
However, the signature verification algorithm has a high computational cost and requires the storage of the public keys of all signers~\cite{boneh2018compact}.
For this reason, this solution is not suitable for the IoT system that we consider.

Finally, in the context of wireless multicasting of data streams, FEC techniques are used to limit the feedback from receivers, in case of packet loss~\cite{nafaa2008forward}.

\begin{figure}[!tb]
\centering
\includegraphics[width=\columnwidth]{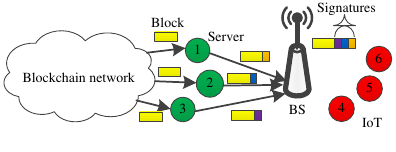}

    \caption{Representation of the system.}
 \label{fig:system_model}
\end{figure}

\section{System model}\label{sec:system}

We assume a set of $V$ servers and a set of $U$ clients, and a BS, as depicted in Fig.~\ref{fig:system_model}.
The servers are observing the updates to the state of a blockchain, by receiving new blocks from the blockchain network.
The block generation process is assumed to be stationary and each new block defines a new period.
Every time the servers receive a valid block, they extract the block header, sign it, and send it to the BS reliably and with a negligible delay.
The block header is signed using public-key cryptography.
To simplify the nomenclature and with a slight abuse of terminology, hereafter we shall refer to the block header as \emph{block}; we also note that the presented analysis is independent from the amount of information in the blocks.

Assuming that all servers are reliable, in block period $t=1,2,\ldots$, the BS receives $V$ copies of the block and $V$ signatures (a pair from each server).
The BS verifies all of them, discarding invalid blocks.

The clients are connected to the BS via wireless links, see the right part of Fig.~\ref{fig:system_model}.
The length of the packet containing a block and a signature packet is $l_b$ and $l_s$ bits, respectively.
We assume that modulation and rate of the BS transmission are fixed and the bit error probability is $P_{\text{bit}}$ for the downlink channels of all clients. Thus, the probability of not receiving a block or signature is given by $p_{e,b} = 1-(1-P_{\text{bit}})^{l_b}$ and $p_{e,s} = 1-(1-P_{\text{bit}})^{l_s}$, respectively.

\begin{figure}[!tb]
\centering
\includegraphics[width=0.8\columnwidth]{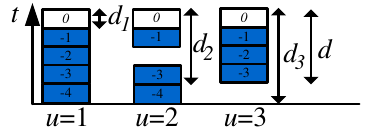}

    \caption{The synchronization process as seen by three clients. The colored blocks are received and authenticated, the empty ones are received by not authenticated.}
 \label{fig:process}
\end{figure}

A client $u \in 1, \ldots,U$ trusts a subset of the servers, for which it knows the public keys. Each client informs the BS of the ID of one of the trusted server during the initialization phase.
When a valid signature is received from the trusted subset of servers, the corresponding block is considered valid, as well as all previously chained blocks.
The client stores the valid blocks and signatures, as shown in Fig.~\ref{fig:process}.
To avoid sending to the IoT devices information that is forked out from the blockchain, servers apply a delay before sending blocks to the BS. The delay, same for all servers, is chosen larger than the maximum duration of a fork. Accordingly, IoT devices never observe the event of fork, which is always resolved on the server side, at the expense of additional information delay.

We denote by $h$ the number of blocks in the blockchain.
Further, we denote by $d_{u}^r$ the difference between $h$ and the last block that $u$ received and that is chained with the genesis block, and by ${d}_{u}^a$ the difference between $h$ and the last block chained with the genesis block for which $u$ received a valid signature.
Note that $d_{u}^a \geq d_{u}^r$.
In the example of Fig.~\ref{fig:process}, $d_{1}^r = 0$, $d_{1}^a = 1$, $d_{2}^r = d_{2}^a = 4$, $d_{3}^r = d_{3}^a = 5$. The maximum tolerated authentication delay, same for all devices, is $d$ block periods. $d$ does not include the delay introduced by servers.

The uplink (UL) channel is assumed reliable within the duration of the block period (that typically lasts several seconds). The UL serves as signal to the BS that the blocks have been received, and to request specific information of interest included in the blockchain state (e.g. ``transactions'' or their ``receipts''~\cite{wood2014ethereum}). However, the latter feature is not analyzed in this work.

\section{Repeat-Authenticate Scheme}\label{sec:schemes}

\begin{figure*}[!tb]
\centering
  {\includegraphics[width=1.7\columnwidth]{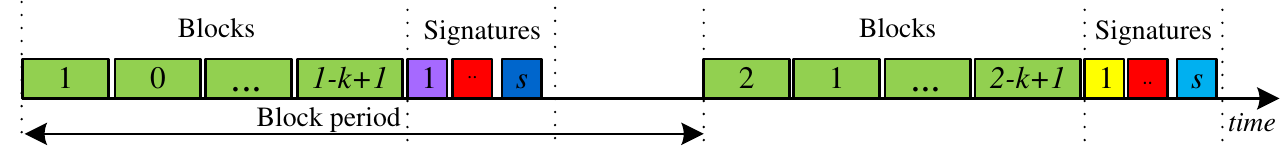}}
    \caption{The information transmitted during block periods $t=1$ and $t=2$.} 
 \label{fig:time}
\end{figure*}

In the proposed scheme, the BS transmits multicast packets containing either a block or a signature, using fixed rate and power. Clients tolerate a delay, hence the BS does not transmit all signatures in all block periods. Instead, in a generic block period the BS sends the most recent $k$ blocks\footnote{This forms a repetition code, which represents the simplest instance of packet-level FEC. Future works may introduce more complex coding schemes.} and $s$ signatures that authenticate the last block among these $k$, see Fig.~\ref{fig:time}.
In each block period, the $s$ signatures to be sent are chosen uniformly at random out of the available $V$ server signatures; this ensures that on average the same number of signatures are transmitted for each server. Each client can possibly receive more trusted signatures per block period by trusting more servers, but the BS is not informed about this.

We assume that $k$, i.e., the number of packets containing blocks in each period, is fixed.
Moreover, the channel resources allocated for the multicast transmissions are fixed to $b$ bits in each block period.
The available resources permit to send
\begin{equation}\label{eq:s}
s = \myfloor*{ \frac{b - k \, l_b}{l_s} }
\end{equation}
signatures in a period, where $l_b$ and $l_s$ are the lengths of a block and a signature, respectively.
It is necessary that $b$ is large enough to allow $s>1$ such that the data stream is authenticated.

The feedback from a client $u$ to the BS consists of a single bit and is transmitted only when $d_{u}^a > d$, where the maximum tolerated authentication delay $d$ is a design parameter. The purpose of the feedback is to trigger a unicast transmission of the last $d$ blocks from the BS to $u$, together with a signature to authenticate all of them, so that $u$ can re-synchronize.
In case of such event, the BS has to allocate extra resources in addition to the $b$ bits, used for multicasting.
According to this mechanism, the BS receives between 0 and $U$ feedback packets in a block period, depending on the synchronization state of the devices.
In the example in Fig.~\ref{fig:process}, client $u=3$ sends the UL packet because $d = 4$ and $d_{3}^a = 5$.
Note that $d$ also provides a upper bound to the value of $k$, as $k \leq d$.

\begin{figure}[!tb]
\centering
\includegraphics[width=0.5\columnwidth]{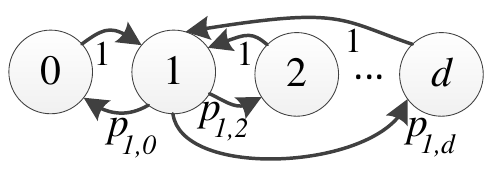}
    \caption{The states of process $x(t)$.}
 \label{fig:sync}
\end{figure}

\begin{figure}[!tb]
\centering
\includegraphics[width=0.9\columnwidth]{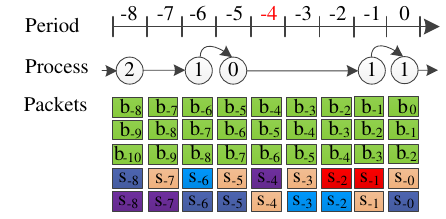}
    \caption{Example of the process, in which $d=4$ and $k=3$. At $t=0$ the client checks if block generated at $t=-4$ is completed. A block and a signature generated at $j$ are indicated as $b_{-j}$ and $s_{-j}$, respectively.}
 \label{fig:ex_analysis}
\end{figure}

\section{Analysis}\label{sec:analys}

In this section, we analyze how to select the parameter $k$ depending on the QoS ($\Phi$) to be offered. The QoS is defined as the long-term average number of clients delayed beyond the deadline $d$ (that is, the number of re-synchronizations triggered):
\begin{equation}\label{eq:phii}
\Phi = \lim_{t \rightarrow \infty} \frac{1}{t}\sum_{\tau=0}^{t} \sum_{u=0}^{U} \mathbbm{1} \{ d_u(\tau) > d \}.
\end{equation}
Ideally $\Phi$ is zero, because there are no delayed clients.

At each period $t$, a client $u\in 1, \ldots,U$ checks if the block generated $d$ periods before (i.e. at $t-d$) has been received \emph{and} authenticated, see Fig.~\ref{fig:ex_analysis}.
Recall that block generated at $t-d$ can be authenticated in two ways: a packet containing its signature is received at $t-d$; or a packet containing a signature is received at $t_j$, $t-d < t_j \leq t$ \emph{and} all blocks between $t-d$ and $t_j$ have been received.
This permits to authenticate the blocks at $t-d$ by leveraging the ledger chain structure. It follows that it is unnecessary to observe the synchronization process at each period, but only when the client may possibly fail. For instance, if the most recent block (generated at $t$) is chained and authenticated, the client will surely not fail within the next $d$ periods.

Based on this key consideration, we introduce the variable $x(t) \in \{ 0, \ldots,d \}$ that is used to track the authentication state the client. When the client may request a unicast re-synchronization, the state is $x(t)=1$. From here, if the block $t-d$ is not authenticated, the client transitions to state $x(t+1)=0$, corresponding to the unicast re-synchronization, at which the last $d+1$ authenticated blocks are sent.
After the client has been re-synchronized, it cannot request a unicast re-synchronization for the next $d$ periods, and hence it spends $d$ periods in state $0$ before going back to state $x(t+d+1) = 1$.
On the other hand, if in state $x(t)=1$ the block $t-d$ has been authenticated by a signature received at $t-d+j$, where $0 \leq j \leq d-1$, the client does not fail. In our process, it corresponds to a transition from $x(t)=1$ to $x(t+1)=j+1$.
The sojourn time in $x(t+1)=j$ is $j$, because the successive $j-1$ blocks are authenticated as well.
From here, the client always returns to $x(t+j-1)=1$.
A possible interpretation of $x(t)$ is that it tracks the \emph{oldest} chained signature found between $t-d$ and $t$. For instance, if a client always receives a chained signature in each period, it would stay in state $x(t)=1 \, \forall t$. If it never receives trusted multicast signatures, it only receives unicast re-synchronizations, looping between $x=0$ and $x=1$. Finally, if the channel is reliable and the client receives a trusted signature every $d$ periods, it loops between states $x=1$ and $x=d$.
In the following text we find the expressions for the transition probabilities $p_{1,j}$ for $0 \leq j \leq d$.

The probability that block at $t-d$ is signed and received is
\begin{equation}
p_{1,1} = p_s \left( 1 - p_{e,b}^k \right),
\end{equation}
where $p_s$ is the probability of receiving the packet (from at least one of the trusted servers) with a signature of the block and $p_{e,b}^k$ the probability of not receiving a packet containing the block in any of the $k$ transmission attempts.
To find $p_s$, we indicate the number of servers trusted by client $u \in 1, \ldots,U$ as $V_u$, $0 < V_u \leq V$. Hence, the probability that it receives at least one signature from a trusted server, in a period, depends on both the packet loss and the probability that the signature of a specific server is sent:
\begin{align}\label{eq:ps}
p_s &= \sum_{j=1}^{\min (V_u, s)} \text{Pr[any of $j$ rx $|$ $j$ sent ] Pr [$j$ sent]} \\
\nonumber &= \sum_{j=1}^{\min (V_u, s)} ( 1 - p_{e,s}^j ) \frac{\binom{V_u}{j} \binom{V - V_u}{s - j} }{\binom{V}{s}}.
\end{align}

\vspace*{0.01in}
Here, $\binom{V_u}{j}$ are the number of combinations for which $j$ signatures, out of the $V_u$ available from trusted servers, are selected to be multicasted; $\binom{V - V_u}{s - j}$ are the number of combinations of the signatures from untrusted servers (there are $V - V_u$), of which $s-j$ are transmitted. Finally, there are $\binom{V}{s}$ combinations of $s$ signatures that can be transmitted among the $V$ available at the BS.

Even if the block at $t-d$ is not signed, it may be authenticated by a block completed between $t-d$ and $t$, if they are chained.
The probability that block at $t-d$ is received and not signed, but authenticated by the $j$-th successive block is found as follows.
Since the first (chained) signature is received after $j$, we must take into account that $j-1$ blocks are received but not signed; we also take into account that the most recent $k-1$ blocks (those for which $d-k+1 < j \leq d$) are transmitted less than $k$ times.
The observations lead to the expression:
\begin{align}\label{eq:exprp}
p_{1, j} =
\begin{cases}
 p_s (1-p_s)^{j-1} (1-p_{e,b}^k)^{j}, & \hspace{-1em}\text{if}\, 1 \leq j \leq d-k+1\\
p_s(1-p_s)^{j-1}(1-p_{e,b}^k)^{d-k+1} & \\
\quad\prod_{i=d-k+2}^{j} (1-p_{e,b}^{d-i+1}),& \hspace{-1em}\text{if}\, d-k+1 < j \leq d.
\end{cases}
\end{align}
The derivation is given in Appendix A. Note the similarity with a geometric distribution, that is ``weighted" by the probability of chaining $j$ blocks.
The probability of failure is found as:
\begin{equation}
p_{1,0} = 1 - \sum_{j=1}^{d} p_{1, j},
\end{equation}
and $p_{j, 1} = 1, \, \forall j \neq 1$ directly follows from the definition of the process.

This concludes the characterization of the transition probabilities of the process of Fig.~\ref{fig:sync}.
We can simplify the Markov chain of Fig.~\ref{fig:sync} by grouping all states $j > 1$ into a state $G$. The new chain has three states $\{0, 1, G\}$ with transition probabilities:
\begin{equation}
\mathbf{P} = 
\begin{pmatrix} 
0 & p_{0,1} & 0 \\
p_{1,0} & p_{1,1} & \sum_{j=2}^d p_{1, j}\\
0 & 1 & 0
\end{pmatrix}
\end{equation}
and stationary probabilities equal to:
\begin{equation}
   \boldsymbol{\pi} = \begin{pmatrix}p_{1,0} & 1 & \sum_{j=2}^d p_{1, j} \end{pmatrix}. 
\end{equation}
The average number of periods spent in states $0$, $1$ and $G$ are $S_0 = d$, $S_1 = 1$ and $S_G = \sum_{j=2}^{d} j$, respectively. We find the average time spent in state $1$ as:
\begin{align}
T &= \frac{S_1}{\sum_{i\in \{0, 1, G\}} S_i \pi_{j}}\\
&= \frac{1}{d \cdot p_{1,0} + 1 \cdot p_{1,1} + \sum_{j=2}^{d} j\cdot p_{1,j}}.
\end{align}

Then, the average number of clients that can potentially fail in a period is $U \cdot \pi_1$, since they follow independent processes. Among them, the average number of clients that actually fail in a period is:
\begin{equation}\label{eq:phi_pro}
\Phi = U \cdot T \cdot p_{1,0}.
\end{equation}

\section{Results}\label{sec:results}

The numerical results are obtained considering block header size of $l_b = 640$ bits (the size of a block header in Bitcoin) and signature size of $l_s = 512$ bits.
The BS allocates $b=8000$ bits per block period for the multicasting.
The maximum tolerated delay is $d=10$ blocks.\footnote{In blockchain applications, it is common to wait several periods, ranging from six in Bitcoin~\cite{nakamoto2008bitcoin} to tens in other blockchain systems~\cite{kraken}, before trusting that the block will not be forked out.}

\begin{figure}[!tb]
\centering
\includegraphics[width=\columnwidth]{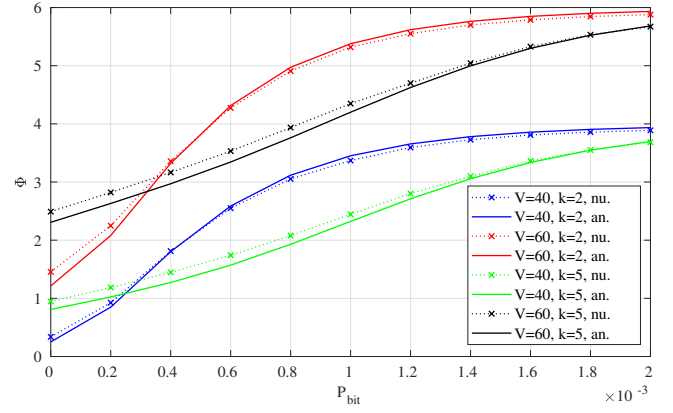}
    \caption{Numerical (nu.) and analytical (an.) values of $\Phi$ for different bit error rates ($P_{\text{bit}}$), repetitions ($k$), and number of clients ($V$).}
 \label{fig:validation_phi}
\end{figure}

First, we study the performance of the scheme, in terms of the the average number of failures, $\Phi$, see Fig.~\ref{fig:validation_phi}, for which  we set $U = V$ and assume that each client trusts a different server.
The figure shows a good match between the numerical and analytical values.
It can be observed that $\Phi$ increases with the bit error probability, $P_{\text{bit}}$, and with the number of servers, as expected.
In the cases when $k$ is 2 and 5, the number of signatures sent in a period are $s=13$ and $s=9$, respectively.
It is worth to note that, if $P_{\text{bit}}$ is low, it is better to reduce the number of repetitions, $k$, in order to accommodate more signatures. However, above a certain threshold value of $P_{\text{bit}}$, the decision is inverted.

\begin{figure}[!tb]
\centering
  \subfloat[]{\includegraphics[width=0.49\columnwidth]{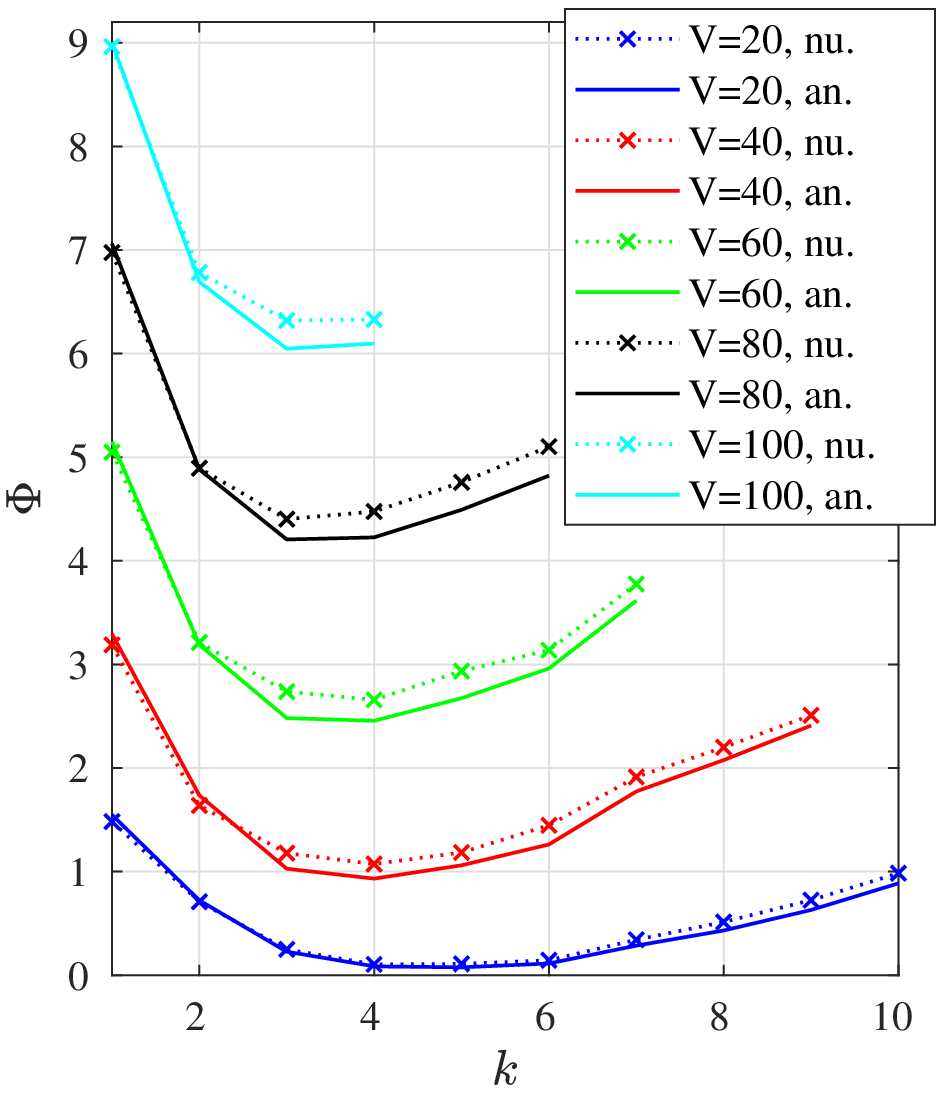}}
   \subfloat[]{ \includegraphics[width=0.49\columnwidth]{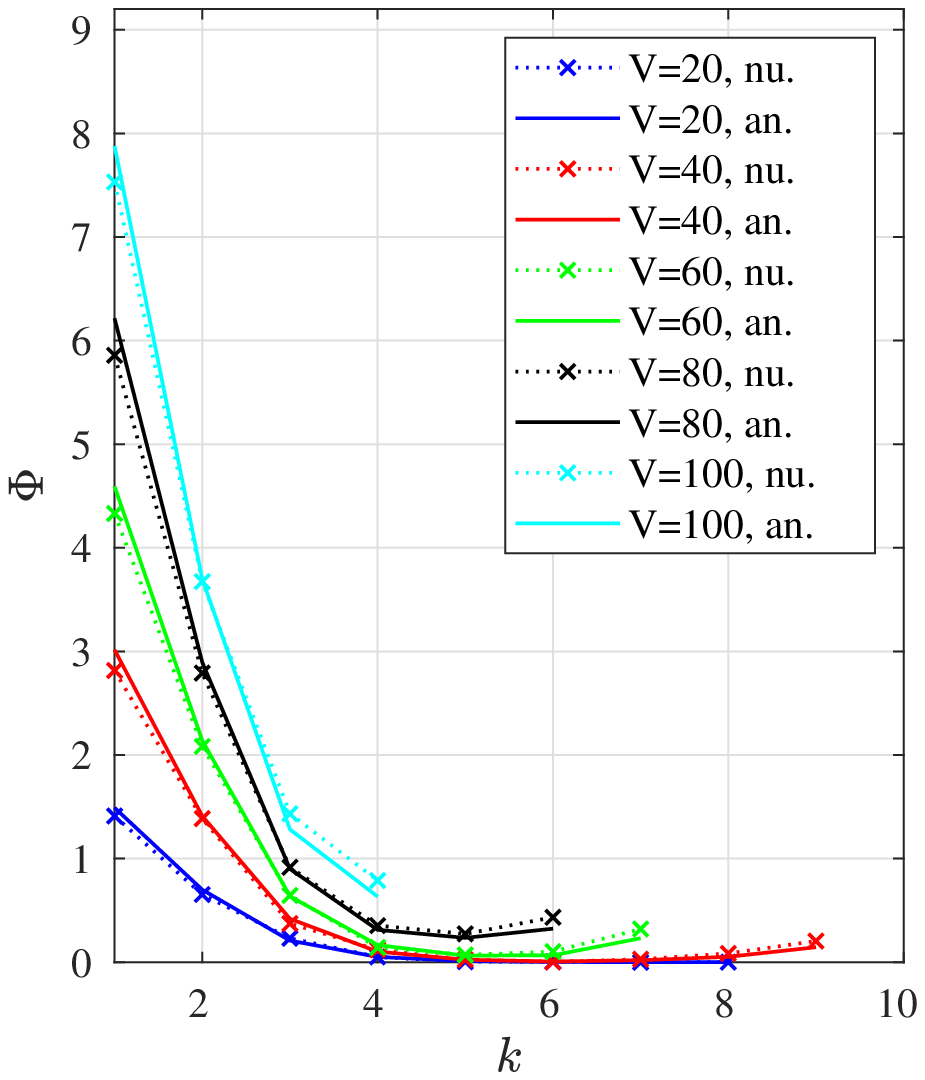}}
    \caption{The average number of failures per block ($\Phi$) for various number of repetitions ($k$), when each client trusts (a) one server and (b) five servers.}
    \label{fig:qos}
\end{figure}

The model also captures the impact of different number of trusted servers per client. Fig.~\ref{fig:qos} reports the average number of failures per block $\Phi$ as function of repetitions $k$ and number of servers trusted by each client. The bit error probability is fixed to $P_{\text{bit}}=4\times 10^{-4}$.
In Fig.~\ref{fig:qos}(a), each client trusts one server. It results that a low number of repetitions are sufficient to increase the reliability. In effect, increasing $k$ reduces the number of signatures per period (see \eqref{eq:s}), then the information is received but not authenticated. 
Fig.~\ref{fig:qos}(b), shows that when a client trusts five (randomly selected) servers, i.e. $V_u=5$, the value of $\Phi$ is drastically decreased, as it disposes of much frequent authentication.
In both graphs, the value of $\Phi$ is only defined in the points where $V/s > d$, that is the condition for which the signature of each server is sent at least once every $d$ periods (on average).

\section{Conclusion}\label{sec:conclusion}

We have introduced the concept of separation of common information, i.e. block headers, and information of interest for single IoT devices, for lightweight blockchain protocols. 
Then, we focused on the efficient transmission of common information and proposed to multicast it, leveraging on the broadcast nature of the wireless channel, as a means to decrease the resources used by the BS, in downlink, and by the IoT devices, for the feedback in UL.
We presented a scheme that provides updates that are timely and authenticated by trusted sources, to a large number of IoT devices. 
The scheme ensures the integrity of the information sent by the BS via digital signatures; however, it is inevitable that the BS can withhold valid blocks or signatures by not transmitting them. The extension in this research direction is part of our current work.

The numerical simulations showed how the performances of the scheme depend on the number of IoT devices, the number of servers that they trust, and the wireless channel quality (bit error rate). 
The scheme finds straightforward application in wireless BS, such as Wi-Fi access points. However, we have only evaluated the parameters of Bitcoin blockchain, in which the size of a digital signature is comparable with the one of a block header. As this is not the case for other blockchains, such as Ethereum, the impact of the ratio between size of the block header and digital signature should be evaluated.
The proposed scheme can, in principle, be extended for the transmission of large block headers or even of the full blocks, if packet fragmentation is taken into account.

In conclusion, our work advocates for further research in improving the connectivity of wireless IoT tailored for blockchain applications.

\section*{Appendix A: Derivation of \eqref{eq:exprp}}

The transitioning from state $1$ to state $j$ happens when the first $j-1$ signatures are not received, and the $j$-th is received. In addition, all the $j$ blocks should be received, to allow the authentication chaining.
We distinguish between two cases.

In the first case, $j \leq d-k+1$, each block is transmitted exactly $k$ times, and we write:
\begin{align}\label{eq:explanation}
    p_{1, j} =& (1-p_s)^{j-1} (1-p_{e,b}^k)^{j-1} p_s (1-p_{e,b}^k)\\ \nonumber
    =& p_s (1-p_s)^{j-1} (1-p_{e,b}^k)^{j}.
\end{align}
Note that, in \eqref{eq:explanation}, we account for $j-1$ unsuccessful transmissions of packets containing signatures; $j-1$ successful transmissions of packets containing blocks (each repeated $k$ times); one successful signature transmission, and successful transmission of the $j$-th block.

In the second case, $j > d-k+1$, we take into account that blocks indexed $d-k+1 < i \leq j$ are transmitted less than $k$ times, because they are recent. For instance, block with index $i = d-k+2$ is transmitted $k-1$ times, block $i =  d-k+3$ is transmitted $k-2$ times. The last block ($j=d$) is transmitted only once.
Based on these observations, we write
\begin{align}
\label{eq:p1}    p_{1, j} =& (1-p_s)^{d-k+1} (1-p_{e,b}^k)^{d-k+1} \cdot \\
\label{eq:p2} & \left( \prod_{i=d-k+2}^{j-1} (1-p_s) (1-p_{e,b}^{d-i+1}) \right) \cdot\\
\label{eq:p3}    & p_s (1-p_{e,b}^{d-j+1}),
\end{align}
where \eqref{eq:p1} takes into account that the first $d-k+1$ blocks are transmitted $k$ times, \eqref{eq:p3} that the $j$-th block is the only one whose signature is successfully transmitted (with probability $p_s$). Finally, \eqref{eq:p2} takes into account the intermediate blocks that are transmitted less than $k$ times. The equation can be re-arranged as:
\begin{equation}\nonumber
    p_{1, j} = p_s(1-p_s)^{j-1}(1-p_{e,b}^k)^{d-k+1} \prod_{i=d-k+2}^{j} (1-p_{e,b}^{d-i+1}).
\end{equation}
The two cases correspond to those of \eqref{eq:exprp}.

\section*{Acknowledgment}

This work has been in part supported the European Research Council (ERC) under the European Union Horizon 2020 research and innovation program (ERC Consolidator Grant Nr. 648382 WILLOW) and Danish Council for Independent Research (Grant Nr. 8022-00284B SEMIOTIC).

\IEEEpeerreviewmaketitle

\nocite{*}
\bibliographystyle{IEEEtran}
\bibliography{refs}

\end{document}